%% file: Paper.tex
\documentclass[aps,eqsecnum,nofootinbib,11pt]{revtex4}

\usepackage{amsmath}
\usepackage{amssymb}
\usepackage{ifpdf}
\usepackage[pdftex]{color,graphicx}
\usepackage{color}
\usepackage[pdftex,bookmarks,colorlinks]{hyperref}
\usepackage[pdftex,bookmarks,colorlinks]{hyperref}
\usepackage{hyperref}
\hypersetup{pdftex}

\begin{document}
\medskip  
\title{Examining  the Field of Static Point Scalar Charge in Schwarzschild Spacetime}
\author{Swapnil Tripathi}
\email{tripathi@uwm.edu}
\affiliation{
Department of Physics, University of Wisconsin-Milwaukee, P.O. Box 413,  
Milwaukee, WI 53201}

\date{\today}

\begin{abstract}

In this paper we examine some aspects of the field of a scalar point charge in curved spacetimes. First we find the closed form solution for the scalar field due to a point charge in Schwarzschild spacetime. Then we expand it locally in powers of r (coordinate distance from the charge)  and compare it to Quinn's local expansion for the field of scalar charge. We show a term by term match, except for a mismatch in one term arising due to an error in a  Riemann tensor term in Quinn's expression. We show the correct expression,  the detailed derivation of which will be subject of a later paper.

\end{abstract}
\pacs{04.25.-g,04.30.-w,04.70.-s,04.70.Bw,04.70.Dy}
\maketitle

\section{Introduction}
\input{Introduction.tex}

\newpage

\section{Notations and Conventions}
\input{Convention.tex}
\section{Field Equation   }
\input{Field.tex}

\newpage

\section{Solution of field equation}
\input{Solution.tex}

\newpage

\section{Quinn's Local Expansion of Scalar Field}
\input{QuinnExp.tex}
\newpage

\section{Conversion of Coordinates}
\input{Coordinates.tex}

\newpage

\section{Comparison}
\input{Comparison.tex}

\section{Conclusion}
\input{Conclusion.tex}
\newpage

\section{Acknowledgments}

It is a pleasure to thank Alan Wiseman for suggesting this problem and providing guidance in solving the problem. I would also like to thank John Friedman for many useful conversations and helpful suggestions. I would like to thank Gonzalo Olmo for providing help in overcoming some mathematical hurdles and Tobias Keidl for proofreading the manuscript and suggesting ways to improve the paper.

%

\include{bib}

\end{document}

%% file: Introduction.tex

The significance of the present paper is in the context of calculation of self-force on particles in curved spacetime. Self-force is in action when a particle feels a force due to its own field. Fields of the particles are distorted by boundary conditions or the structure of the spacetime and act back on the particle to create this self-force. A simple example of Newtonian gravitational self-force is seen when a small satellite goes around a larger central star or planet. The satellite does not orbit around the center of the larger object, but around a common center of mass of both bodies. The radius of this orbit is smaller than the distance between the centers of two masses. The mass of the satellite produces a perturbation in the gravitational field of the central object and follows a geodesic in this altered field. So the gravitational field of the body itself affects the path it follows in spacetime. This is called self-force.

This perspective involves action at a distance. Another way to look this is that the moving particle creates a gravitational ripple in the spacetime which travels outward. This ripple or wave hits other bumps in spacetime like those created by other massive objects or boundary conditions and then comes back to interact with the original particle again causing a deflection in its path. The failure of HuygensÕ principle produces a tail in the spacetime behind the leading edge of the wave that creates self-force.

Understanding of this phenomenon is important for several reasons. At the most basic level calculating self-force is a way of understanding the behavior of classical fields in curved spacetimes. It is also necessary for getting a better understanding of the two-body problem in general relativity. Explained earlier was the Newtonian correction to the trajectory due to the objects own field. Post Newtonian effects due to the self-field need to be taken into account to more closely model the actual trajectory of the objects in the spacetime.This increases our understanding of the radiation reaction problem.

In the current paper we examine the field of a scalar charge in Schwarzschild spacetime. The field of scalar charge can be used to calculate the self-force of a scalar particle using the Quinn-Wald Axioms.

%% file: Convention.tex
The signature of the metric $g_{\mu\nu}$ is {-+++}. Here commas denote partial differentiation and the indices run from 0 to 3.  

The Riemann Tensor is given by
\begin{eqnarray}
R^{ d}_{cab}=\Gamma^{d}_{cb,a}-\Gamma^{d}_{ca,b}+\Gamma^{e}_{bc}\Gamma^{d}_{ae}-\Gamma^{d}_{be}\Gamma^{e}_{bc} .
\end{eqnarray}
It should be noted that this differs from the convention of Dewitt and Brehme's\cite{dewittbrehme} Riemann Tensor:
\begin{eqnarray}
\mathcal{R}^{\quad d}_{abc}=\Gamma^{d}_{cb,a}-\Gamma^{d}_{ca,b}+\Gamma^{e}_{cb}\Gamma^{d}_{ea}-\Gamma^{d}_{eb}\Gamma^{e}_{cb} ; \qquad (DB).
\end{eqnarray}
We compare the two expressions and find,
\begin{eqnarray}
\underset{DB}{\mathcal{R}^{\quad d}_{abc}}=-  {R^{\quad d}_{abc}}.
\end{eqnarray}

%% file: Field.tex

In this section we study the radiation
reaction problem associated to a scalar particle. Scalar charges are
idealized objects which have been extensively used in the literature
to gain insight on different physical aspects of the radiation
reaction problem while, at the same time,  reducing the technical
complexity that real particles and fields would involve. Since a
standardized convention does not exist in the literature, our first
task will be to determine whether our scalar charges attract or repel.
We thus define our action as follows,
\begin{eqnarray}
\mathcal{A}&=&\int \Big [\mathcal{L}_\textrm{field}+\mathcal{L}_\textrm{interaction}+\mathcal{L}_\textrm{particle} \Big]\sqrt{-g} d^{4}x
\nonumber\\
&=& \int \Big [-\frac {1}{8\pi}g^{\alpha'\beta'}\nabla_{\alpha'}\Phi\nabla_{\beta'}\Phi+\rho\Phi+\frac12 \int m_{0}g_{\alpha'\beta'}u^{\alpha'}u^{\beta'}\delta^{4}(x-z(\tau)) d\tau\Big]\sqrt{-g} d^{4}x.
\end{eqnarray}
Here, $\Phi$ is the scalar field, $\rho$ is the scalar charge density and $\sqrt{-g}$ is the determinant of the covariant components of the metric. The term $\mathcal{L}_{field}$ represents the action associated with the field of the scalar particle.  The term $\mathcal{L}_{particle}$ represents the contribution to the action from the bare mass $m_{0}$ of the particle, and $\mathcal{L}_{interaction}$  is the  contribution from interaction of charge to the field.

Varying the action with respect to the field and making it invariant gives the field equation. The
Euler-Lagrange equations lead to
\begin{eqnarray}
 -\frac{1}{8\pi} \frac{\partial}{\partial x^{\alpha'}} 2\Phi_{\cdot \alpha'}=\rho.
\end{eqnarray}

Therefore, our field equation is
\begin{eqnarray}\label{field} 
\Box \Phi=-4\pi\rho.
\end{eqnarray}

Note the ``-''  sign in the equation.  This represents the scalar particles repelling each other. Work must be done to assemble the charges.  Therefore additional energy or ``mass'' is stored in the particle when one assembles a set of scalar charges.  This is important for renormalizing the mass.  For a static charge in flat spacetime at the origin, the above field equation can be written as
\begin{eqnarray}
\nabla^{2}\Phi=-4\pi q \delta^{3}(x).
\end{eqnarray}

The solution to this equation is
\begin{eqnarray}
\Phi=\frac q r.
\end{eqnarray}

The energy of interaction for another charge $q$ in this field will be 
\begin{eqnarray}
V_{12}=q\Phi=\frac{q^{2}}{r},
\end{eqnarray}
which is positive.  This implies a positive energy is added to the system when assembling a ball of charge.  This results in a contribution to the renormalization of the mass.  We define
\begin{eqnarray}
m_{ren}=m_{0}+\frac {q^{2}}{2\epsilon}.
\end{eqnarray}

A different sign convention, say the one used by Shapiro-Teukolsky \cite{shapiro} yields the field equation 
\begin{eqnarray}
\Box \Phi=4\pi\rho,
\end{eqnarray}
which implies that like scalar charges attract and therefore a set of charge has negative energy.  The mass renormalization term \ has the opposite sign in this case.

%% file: Solution.tex
We now proceed to solve
equation Eq.~[\ref{field}]  which is analogous to the field equation for electric charges. The source $\rho$ for the field is a point scalar charge given by
\begin{eqnarray}
\rho(t,\textbf{x}) = q \int_{-\infty}^{\infty} (\frac{1}{\sqrt{-g}}) \delta^4(x^\alpha-b^\alpha(\tau)) d\tau,
\end{eqnarray}
where $ \delta^4(x^\alpha-b^\alpha(\tau))$ is the four dimensional Dirac delta function and $b^\alpha(\tau)$ is the coordinate of the scalar charge.  This equation is solved in detail for a static scalar charge in Wiseman\cite{wiseman}.  We find the  field \footnote[1]{We do not repeat the calculation here;  we only use its results with an important note.  Whereas Wiseman uses a convention in which like scalar charges attract, we use a convention where like scalar charges repel. The difference in convention manifests itself as a positive or negative sign respectively in front of the source term in field equation.} to be 
\begin{eqnarray}
 \phi= q  \sqrt{\frac{b_{h}-M}{b_{h}+M}} \frac{1}{\sqrt {r_{h}^{2}-2r_{h}b_{h}\cos{\theta} +b_{h}^{2}-M^{2}\sin^{2}{\theta}}}.
\end{eqnarray}
Wiseman's result is analogous to an old result due to
Copson\cite{copson}, modified by Linet\cite{linet}  which gives a closed-form expression for the electrostatic
potential of a fixed charge residing in Schwarzschild spacetime. 
\begin{figure}[!h]
\begin{center}
\includegraphics{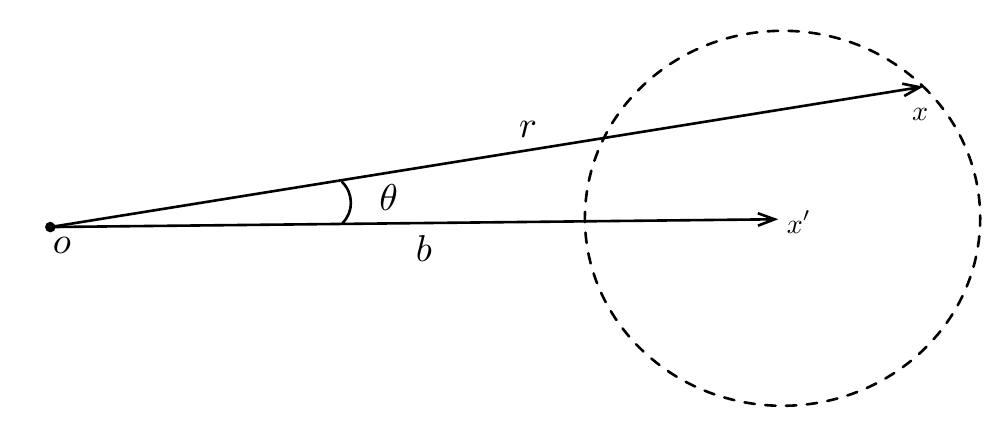}
\caption{Schwarzschild Coordinate}
\label{default}
\end{center}
\end{figure}
In the figure, $x^{\prime}$ marks the location of the charge and x marks the field point. The subscript h stands for harmonic coordinates. Harmonic coordinates are given by, $r_h=r_s-b$, where $r_s$ is the radial coordinate in Schwarzschild coordinates.
This is a particular solution of the inhomogeneous field equation and the field satisfies the boundary conditions:  it is regular at both infinity and the horizon.The solutions for the homogeneous part of the field equation Eq.~[\ref{field}] are hypergeometric functions\cite{saul}, Legendre functions in particular\cite{wiseman}.  These are not well behaved at both the horizon and infinity.   This implies that there exist no smooth solutions which can be added to the particular solution to give a set of different solutions.  Therefore the above particular solution is the unique solution for the field of ``static" scalar charge in Schwarzschild spacetime.  Note that this argument is only applicable for static solutions of scalar and electromagnetic fields in Schwarzschild spacetimes. For further discussion of this the reader is referred to Ref.~\cite{saul} and Ref.~\cite{wiseman}.

This has another important consequence for the ``tail" term in Quinn's expansion for the scalar field which we discuss here briefly. The tail term is the smooth part of the scalar field\cite{poisson} and it satisfies the homogeneous equation.  Since no smooth solutions exist for static scalar charge in Schwarzschild spacetime, this implies that ``tail" term vanishes for this case.

By definition the tail of the Green's function produces self-force.  There are no self-force effects in the static  scalar Schwarzschild case from the Abraham-Lorentz-Dirac radiation reaction (ALD) or coupling to Ricci curvature.  The total self-force has been shown to vanish\cite{wiseman} which indicates that the tail of Green's function vanishes here.

%% file: QuinnExp.tex
Quinn\cite{quinn}  calculated a local expansion of the field of a scalar charge moving on an arbitrary trajectory in a general spacetime.  Quinn's result for the field and its gradient are:
\begin{eqnarray}
 \phi_{\pm}(x)=q \Bigg ( \frac1r - \frac12 a^a \hat r_a \Bigg ) \pm  \lim_{\epsilon\rightarrow 0} \int^{\pm\infty}_{\tau_{\pm} \pm \epsilon} G_{\pm}(x,z(\tau)) d\tau + O(r),
\end{eqnarray}
and
\begin{eqnarray} \label{eq:grad2}
\nabla_{a'} \phi_{\pm}=q \bar g_{a'a} \Bigg (&-&\frac{\hat{r}^{a}}{r^{2}} - \frac12 \frac{a^{a}}{r}+\frac12 \frac {(a^b\hat r_b)} {r} \hat{r}^{a} -\frac{3}{8} (a^b\hat r_b)^2 \hat{r}^{a} +\frac{3}{4} (a^b\hat r_b) a^a 
\nonumber\\
 &-&\frac16 R_{bdce} u^b u^c  \hat{r}^{d} \hat{r}^{e} \hat{r}^{a} -\frac18 a^2  \hat{r}^{a} -\frac{1}{12} R_{bc}  \hat{r}^{b}  \hat{r}^{c}  \hat{r}^{a} + \frac12 (\dot a^b  \hat{r}_{b}) u^a
 \nonumber\\
  &+&\frac{1}{12} R_{bc}  u^{b}  u^{c}  \hat{r}^{a} + \frac{1}{6} R_{bc}  u^{b}  \hat{r}^{c} u^{a} + \frac13 R^a_{\  cbd} u^b u^c \hat r^d \pm \frac13 a^2 u^a \mp \frac13 \dot a^a
 \nonumber\\
&\mp& \frac{1}{6} R_{bc}  u^{b}  u^{c}  u^{a} +R^{ab} \hat r_b -\frac{1}{12} R \hat r^a \mp \frac16 R^{ab} u^b \pm \frac{1}{12} R u^a \Bigg )
 \nonumber\\
&\pm& \lim_{\epsilon\rightarrow 0} \int^{\pm\infty}_{\tau_{\pm} \pm \epsilon} \nabla_a G_{\pm}(x,z(\tau)) d\tau + O(r).
\end{eqnarray}


In later sections of this paper we will do a term by term comparison of these expansions with expansions of closed form solutions for static scalar charge in Schwarzschild spacetime. The above expressions simplify in the following ways:

\begin{itemize}
\renewcommand{\labelitemi}{$\bullet$}
\item {The ``tail of Green's function" (the integral of Green's function over past history) vanishes for static scalar charge in Schwarzschild spacetime.}
\item Ricci scalar and Ricci tensors  vanish in Schwarzschild Spacetime.
\item For a static particle there are no separate advanced/retarded field. There will be only one field and time asymmetric terms will vanish.
\item We only look at field point lying on the axis joining the origin to the particle.We call this line  $z$-axis.
\item The acceleration (static in Schwarzschild) is $a^r=\displaystyle\frac {m}{b^2}$ where m is the mass of the central black hole and b is the radial coordinate of the particle along $z$-axis $a^2=|a^\alpha a_\alpha|= \displaystyle\frac {m^2}{b^4 (1-2m/b)}$.
\end{itemize}

On applying these simplifications the field is

\begin{eqnarray} \label{eq:field}
 \phi(x)=q \Bigg ( \frac1r - \frac {m}{b^2 \sqrt{1-2m/b}}\Bigg ) + O(r),
\end{eqnarray}
and the gradient is
\begin{eqnarray}
\nabla_{a'} \phi=q \bar g_{a'a} \Bigg (&-&\frac{\hat{r}^{a}}{r^{2}}  -\frac{3}{8} (a^b\hat r_b)^2 \hat{r}^{a} +\frac{3}{4} (a^b\hat r_b) a^a -\frac18 a^2  \hat{r}^{a} 
\nonumber\\
 &-&\frac16 R_{bdce} u^b u^c  \hat{r}^{d} \hat{r}^{e} \hat{r}^{a}  + \frac13 R^a_{\  cbd} u^b u^c \hat r^d \Bigg )+ O(r).
\end{eqnarray}
\begin{itemize}
\renewcommand{\labelitemi}{$\bullet$}
\item The two acceleration terms in the order of (1/r) cancel with each other for acceleration directed along the $z$-axis.
\item $\dot a = 0$.
\end{itemize}
The $\bar g_{a'a} $ is bivector of geodetic parallel transport\footnote{For detailed discussion of bitensor fomalism, the reader is directed to Ref.~\cite{dewittbrehme}}.
For purposes of current calculation it is sufficient to know that 
\begin{eqnarray}
\bar g_{a'a} = \delta_{a'a} + \frac16 r^2 R_{\alpha\gamma\beta\delta} \hat r^\gamma \hat r^\delta.
\end{eqnarray}
When it is dotted into the bracket, $\delta_{a^\prime a}$ has lowers the indices and changes them from $a$  to $a^\prime$($z$ to $z^\prime$) for this case.  In this notation the primed indices represent the field point and the unprimed indices represent the source point. The Riemann term only interacts with the $-\displaystyle \frac{\hat{r}^{a}}{r^{2}}$ term to lower the order in r, but the Riemann component vanishes due to symmetry by contracting with radial coordinate three times.  The expression for gradient further simplifies to
\begin{eqnarray}
\nabla_{z'} \phi=q  \Bigg (-\frac{\hat r_z}{r^2}  -\frac{3}{8} a^2 \hat{r}_z +\frac{3}{4} a^2 \hat r_z -\frac18 a^2  \hat{r}_{z} 
 -\frac16 R_{0d0e} u^0 u^0  \hat{r}^{d} \hat{r}^{e} \hat{r}_{z}  + \frac13 R_{zooz} u^0 u^0 \hat r^z \Bigg )+ O(r).
\end{eqnarray}

Now all the quantities on RHS have unprimed indices meaning they are calculated at the source point. The gradient is naturally calculated at the field point. The ``geodetic bivector of parallel transport" transports the quantities being calculated at the source to the field point. After acting with the $\bar g_{a'a}$ on the bracketed term, we find that on the $z$-axis, the calculation of the gradient using quantities at source point will give us the same value as at the field point.
\begin{eqnarray}
\nabla_{z'} \phi=q  \Bigg (&-&\frac{\hat r_z}{r^2} +\frac{1}{4} a^2 \hat r_z  -3\frac16 R_{0z0z} u^0 u^0  \hat{r}^{z} \hat{r}^{z} \hat{r}_{z}  + \frac13 R_{zooz} u^0 u^0 \hat r^z \Bigg )+ O(r).
\end{eqnarray}

The Riemann Normal Coordinates (RNC) are defined at the the source point with the spatial coordinates orthogonal to the worldline .

\begin{eqnarray}
\nabla_{z'} \phi=q  \Bigg (&-&\frac{\hat r_z}{r^2} +\frac{1}{4} a^2 \hat r_z  -3\frac16 R_{0101} u^0 u^0  \hat{r}^{z} \hat{r}^{z} \hat{r}_{z}  + \frac13 R_{1001} u^0 u^0 \hat r^z \Bigg )+ O(r).
\end{eqnarray}
Therefore the $z$-component of the gradient is
\begin{eqnarray}
\nabla_{z'} \phi=q  \Bigg (&-&\frac{1}{r^2} +\frac{1}{4} a^2    -\frac12 R_{0101}  - \frac13 R_{0101}  \Bigg )+ O(r).
\end{eqnarray}
The $R_{0101} $ component of the Riemann in RNC is $\displaystyle\frac{2m}{b^3}$\cite{zhao}.  
Finally, we find
\begin{eqnarray} \label{eq:grad1}
\nabla_{z'} \phi=q  \Bigg (&-&\frac{1}{r^2} +\frac14 \frac {m^2}{b^4 (1-2m/b)} - \frac53 \frac{m}{b^3}  \Bigg )+ O(r).
\end{eqnarray}
Note the last term has an incorrect coefficient arising from an error in Quinn's expression for the scalar field. This error results from the sign of $\frac16 R_{bdce} u^b u^c  \hat{r}^{d} \hat{r}^{e} \hat{r}^{a} $.  When this is corrected, the final form of the expression is
\begin{eqnarray}\label{eq:grad2}
\nabla_{z'} \phi=q  \Bigg (&-&\frac{1}{r^2} +\frac14 \frac {m^2}{b^4 (1-2m/b)}+ \frac13 \frac{m}{b^3}   \Bigg )+ O(r),
\end{eqnarray} 
which matches with our expansion obtained from the particular solution.  It is important to emphasize that this error does not affect the final self-force result in Quinn's paper because this term vanishes by the angular averaging in his prescription for the self-force calculation.

%% file: Coordinates.tex
Now we can compare the expression obtained from Quinn's expansion with the field we calculated in the earlier section. We need to express both expressions in equivalent coordinates.  Quinn's expansion is done in RNC centered at the source point whereas our field is expressed in Schwarzschild coordinates with origin located at the center of Black Hole.

We will express Wiseman's field in RNCs of the Quinn's expansion to make a comparison.  We will first show that a coordinate distance between source and field point along the $z$-axis in RNCs of Quinn's expansion is essentially the proper distance. Therefore, expressing Wiseman's scalar field as an expansion in $\triangle s$, the proper distance, is equivalent to expressing it in RNCs.  The next step will be to express  $\triangle r =r_s-b_s$, the coordinate distance between the source and field point Schwarzschild coordinates, in terms of $\triangle s$.
\subsection{Riemann Normal Coordinates}

The metric for Riemann Normal coordinates is given by
\begin{eqnarray}
g_{00}&=&-1+\frac13R_{0l0m} x^l x^m,
\\
g_{0i}&=&\frac23R_{0lim} x^l x^m,
\\
g_{ij}&=&\delta_{ij}+\frac13R_{iljm} x^l x^m.
\end{eqnarray}

\begin{figure}[!h]
\begin{center}
\includegraphics{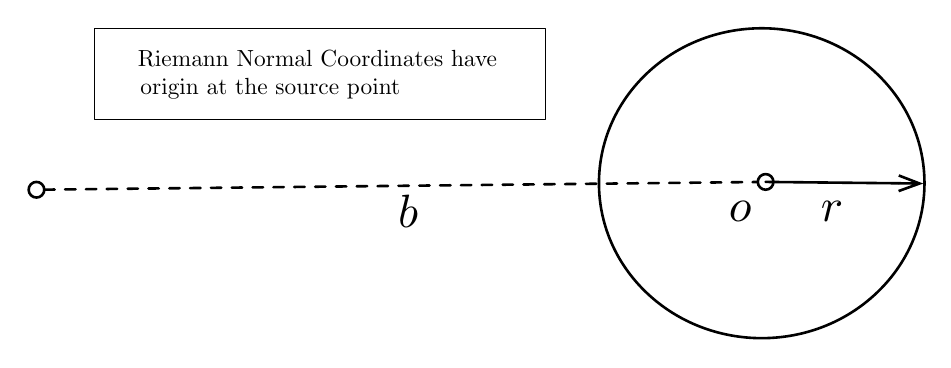}
\caption{Riemann Normal Coordinate }
\label{default}
\end{center}
\end{figure}

For a coordinate distance along z axis, we have
\begin{eqnarray}
ds^2&=&g_{zz} dr^2,
\\
ds&=&\big(\sqrt{ 1+R_{zzzz} z z}\big) dr,
\\
ds&=&dr,
\end{eqnarray}
since Riemann component vanishes by symmetry.  Coordinate distance along $z$-axis $ ``r"$ in Quinn's expression is the same as $\triangle s$  the proper distance.
\subsection{Schwarzschild coordinates}

The Schwarzschild metric is
\begin{eqnarray}
-d\tau^2=ds^2=-(1-\frac{2m}{r_s})dt^2 +(1-\displaystyle\frac{2m}{r_s})^{-1} dr_s^2 + r^2( d\theta^2+ \sin(\theta)^2 d\Phi^2)
\end{eqnarray}
Here we take $ z= r_{s}-b_s$ the coordinate distance between $x$ and $x'$.  A variation in $z$ is related to the change in proper distance by
\begin{eqnarray}
ds= \frac{1}{\sqrt{1-\frac{2m}{b_s+z}}} dz.
\end{eqnarray}

From here onwards we drop the subscript ``s'' in $b_s$  to avoid clutter.  Expanding in powers of z and integrating we find the proper distance:
\begin{eqnarray}
\triangle s=\frac{z}{\sqrt{1-\frac{2m}{b}}} -\frac{mz^2}{2b^2 (1-\frac{2m}{b})^\frac{3}{2}} + \frac{mz^3}{12b^3 (1-\frac{2m}{b})^\frac{3}{2}}+ \frac{mz^3}{4b^3 (1-\frac{2m}{b})^\frac{5}{2}} +O(z^4)
\end{eqnarray}

Inverting the above equation we find
\begin{eqnarray}
z= \sqrt{1-2m/b} \Bigg(\triangle s + \frac {m} {2b^2 \sqrt{1-2m/b}} {\triangle s}^2 + \frac {m} {3b^3} {\triangle s}^3\Bigg) +O({\triangle s}^4).
\end{eqnarray}

From previous section, we know  ``r'' in Quinn's expansion is equal to  $\triangle s$.  Then
\begin{eqnarray} \label{eq:conv1}
z= \sqrt{1-2m/b} \Bigg(r+ \frac {m} {2b^2 \sqrt{1-2m/b}} {r}^2 + \frac {m} {3b^3} {r}^3\Bigg) +O({r}^4),
\end{eqnarray}
can be used for conversion of coordinates.   Also 
\begin{eqnarray}  \label{eq:conv2}
\frac{dz}{dr}= \sqrt{1-2m/b} \Bigg(1+ \frac {m} {b^2 \sqrt{1-2m/b}} r + \frac {m} {b^3} {r}^2\Bigg) +O({r}^3).
\end{eqnarray}





%% file: Comparison.tex
We start with Wiseman's  field for static scalar charge in Schwarzschild spacetime,
\begin{eqnarray}
 \phi= q  \sqrt{\frac{b_{h}-m}{b_{h}+m}} \frac{1}{\sqrt {r_{h}^{2}-2r_{h}b_{h}\cos{\theta} +b_{h}^{2}-m^{2}\sin^{2}{\theta}}},
\end{eqnarray}
where $b_h$ is the $z$-coordinate of the scalar charge in harmonic coordinates. We use $``m"$ instead of $``M"$ for the mass of black hole to keep a consistent notation.  In this equation we define $r_{h}\equiv r_{s}-m$, where $r_{s}$ is the field point in Schwarzschild coordinates and $r_{h}$ is field point in harmonic coordinates.   We restrict our attention to a static scalar charge fixed at $\hat{b}=b \hat z$ .

The field point is located on the $z$-axis such that  $r_s>b_s$. On expressing the field in Schwarzschild coordinates  and setting it on the  z axis  ($\theta  = 0$, so $\cos\theta  = 1$ and  $\sin\theta  = 0$), we find
\begin{eqnarray}
\phi_{axis}=q\sqrt{1-\frac{2m}{b_s}} \frac{1}{|r_{s}-b_s|}.
\end{eqnarray}
Here $b_s$ is the z coordinate of the scalar charge in Schwarzschild coordinates. Recall $z=r_s-b_s$. Again we drop the subscript `s'  from $b_s$ to avoid clutter.  Then
\begin{eqnarray}
\phi_{axis}=q\sqrt{1-\frac{2m}{b}} \frac{1}{z}.
\end{eqnarray}

Converting to RNC using Eqn.~(\ref{eq:conv1}), we find 
\begin{eqnarray} 
 \phi_{axis}&=&q \sqrt{1-\frac{2m}{b}}\Bigg ( \frac{1}{r \sqrt{1-2m/b}} - \frac {m}{b^2 ({1-2m/b})}\Bigg ) + O(r),
\nonumber \\
  &=&q\Bigg ( \frac{1}{r } - \frac {m}{b^2 \sqrt{1-2m/b}}\Bigg ) + O(r),
\end{eqnarray}
which is equivalent to Eqn.~(\ref{eq:field}).

When finding the gradient of the field, we include the next higher order term in the expression for field and then take a derivative with respect to $r$ (the radial coordinate in RNC).  This gives
\begin{eqnarray}
\nabla_r \phi_{axis}&=&q  \frac{d}{dr} \Bigg ( \frac{1}{r } - \frac {m}{b^2 \sqrt{1-2m/b}}+ \bigg(\frac14 \frac {m^2}{b^4 (1-2m/b)}+ \frac13 \frac{m}{b^3}\bigg)r + O(r^2) \Bigg ) 
\nonumber\\
&=&q \Bigg ( -\frac{1}{r^2 } + \frac14 \frac {m^2}{b^4 (1-2m/b)}+ \frac13 \frac{m}{b^3} \Bigg )+ O(r) 
\end{eqnarray}

This result is identical to  Eqn.~(\ref{eq:grad2}) which was obtained by correcting a sign error in the term involving Riemann tensor. The sign in front of the term $\displaystyle \frac16 R_{bdce} u^b u^c  \hat{r}^{d} \hat{r}^{e} \hat{r}^{a} $ in Quinn's Eqn.(\ref{eq:grad2}) should have a ``+'' instead of  ``-''.

%% file: Conclusion.tex
We have shown the equivalence of expression (with correction) calculated by Quinn for the field of scalar charge in Schwarzschild spacetime  with the analytical form for the field. Quinn's local expansion for the field was calculated using Dewitt-Brheme's world function. We have shown that Quinn's expression contained a minor error in the sign of one of the Riemann terms. It went unnoticed in his calculation because the term was averaged away in his prescription of self-force calculation and so his result for self-force was unaffected and was correct.  In this paper we show the error and the correction. The derivation of the correct term will be subject for another paper.
 

%% file: bib.tex

%% file: Paper.bbl
\begin{thebibliography}{99}
\bibitem{coordinates} We use units in which $G=c=1$. Our conventions
are those of Misner, Thorne and Wheeler \cite{mtw}. 
In particular, only Greek indicies are summed.
The line element is  the 
Schwarzschild metric in standard coordinates $ds^2 = -
(1-2M/r)dt^2  + (1-2M/r)^{-1} dr^2 + r^2 (d\theta^2 +\sin^2
\theta d\phi^2)$.  The point charge is placed on the $z$-axis, 
and the angle $\theta$ is also measured from the $z$-axis.
In comparing our results with, say
\cite{wiseman}, one has to convert to a harmonic radial coordinate,
$r_h = r-M$.
\end{thebibliography}
